\documentclass{appolb}
\usepackage{graphicx}
\usepackage{graphicx}
\usepackage{multirow}
\usepackage{amsmath,amssymb,amsfonts}%
\usepackage{lineno}
\usepackage{mathrsfs}%
\usepackage[title]{appendix}%
\usepackage{textcomp}%
\usepackage{manyfoot}%
\usepackage{booktabs}%
\usepackage{algorithm}%
\usepackage{algorithmicx}%
\usepackage{tabularray}
\usepackage{nicematrix}  
\usepackage{soul} 
\usepackage{algpseudocode}%
\usepackage{listings}%
\usepackage{adjustbox}
\usepackage{subcaption}
\usepackage{makecell}
\usepackage{rotating}

\begin{document}
\title{Deep Learning Approaches for BSM Physics: Evaluating DNN and GNN Performance in Particle Collision Event Classification}

\author{Ali CELIK
\address{Department of Physics, Burdur Mehmet Akif Ersoy University}
\\[3mm]
}
\maketitle
\begin{abstract}
Detecting Beyond Standard Model (BSM) signals in high-energy particle collisions presents significant challenges due to complex data and the need to differentiate rare signal events from Standard Model (SM) backgrounds. This study investigates the efficacy of deep learning models, specifically Deep Neural Networks (DNNs) and Graph Neural Networks (GNNs), in classifying particle collision events as either BSM signal or background. The research utilized a dataset comprising 214,000 SM background and 10,755 BSM events. To address class imbalance, an undersampling method was employed, resulting in balanced classes. Three models were developed and compared: a DNN and two GNN variants with different graph construction methods. All models demonstrated high performance, achieving Area Under the Receiver Operating Characteristic curve (AUC) values exceeding $94\%$. While the DNN model slightly outperformed GNNs across various metrics, both GNN approaches showed comparable results despite different graph structures. The GNNs' ability to explicitly capture inter-particle relationships within events highlights their potential for BSM signal detection.
\end{abstract}
  
\section{Introduction}
The Standard Model of particle physics \cite{salam1968weak,weinberg1967model} has been extensively tested by the CMS and ATLAS experiments, with experimental data confirming its remarkable success. However, despite its achievements, the SM cannot explain some phenomena, which leads us to search for beyond the Standard Model theories. At CERN, the CMS and ATLAS collaborations have conducted searches for dark matter candidates, one of the primary phenomena that the SM cannot explain. While these searches have not yielded positive results thus far, they have placed mass limits on the sought-after particles. In one of these studies \cite{atlas2022search}, in the scenario where a pair of gluinos is produced and these gluinos decay to the LSP via off-shell stop or sbottom, gluino masses below 2.44 TeV and 2.35 TeV, respectively have been excluded at $95\%$ CL for the case of a massless lightest supersymmetric particle called LSP ($\tilde{\chi}_1^{0}$).

Although the production cross-section of colored particles in hadron colliders is higher than that of electroweakinos, the limits placed on these particles are also higher compared to electroweakinos. However, the mass limits placed on electroweakinos are still within the reach of CMS or ATLAS experiments. In a study \cite{aad2023searches_squark_1550} examining the direct production mechanism of chargino-neutralino ($\tilde{\chi}_1^{\pm}$/$\tilde{\chi}_2^{0}$), a mass limit of 820 GeV was placed on the  ($\tilde{\chi}_1^{\pm}$/$\tilde{\chi}_2^{0}$) for the case of a massless $\tilde{\chi}_1^{0}$. A separate study \cite{aad_search_2020}, which investigated slepton ($\tilde{l}$) pair production, where each slepton decays into a lepton and an LSP, excluded slepton masses up to 700 GeV.

In another study \cite{aad2021search-5} conducted by the ATLAS collaboration, which investigated SUSY scenarios with R-parity violation, a comprehensive analysis was performed using 139 $fb^{-1}$ of data. This research examined various production mechanisms, including gluino pair production, stop pair production, chargino-neutralino production, and neutralino pair production. The results of this study provided significant constraints on SUSY particle masses. For scenarios with high LSP mass, gluino masses up to 2.4 TeV were excluded. In cases with low LSP mass, the exclusion limit for gluino masses reached up to 2 TeV. Notably, the study also set stringent limits on the top squark mass, excluding it up to 1.35 TeV, further constraining the parameter space for supersymmetric models. Furthermore, the study provided important insights into the mass limits of the LSP itself. For higgsino LSP scenarios, where the LSP is assumed to be mass-degenerate with the second-lightest neutralino, LSP masses up to 320 GeV were excluded. In the case of wino LSP, where the LSP is assumed to be mass degenerate with the lightest chargino, the exclusion limit for LSP masses was extended up to 365 GeV.

Given the challenges in detecting electroweakinos due to their lower production rates, researchers have turned to advanced analytical techniques to enhance their search capabilities. Therefore, different research groups need to utilize various methods to increase acceptance in studies addressing this production mechanism. One of these methods is to utilize machine learning algorithms. Machine learning algorithms can be successful in capturing linear or non-linear relationships between features in scientific fields like particle physics, where the collected data is multi-dimensional and the classical cut-and-count method may not yield sufficiently good results. A study conducted by \cite{anderssen_thesis} reproduced the results of studies \cite{aad_search_2020,aaboud_search_2018} performed by the ATLAS group, using machine learning algorithms, which addressed direct slepton production or chargino/neutralino production and their decay into different final states via $\tilde{l}/W^{\pm}$ boson. The obtained results showed that the machine learning method enhanced the outcomes. Machine learning algorithms are not only used in physics analyses but are also widely utilized in the reconstruction of physics objects and in the jet-flavor tagging stage in CMS and ATLAS experiments \cite{cms2022identification_tau,atlas2020deepsets,cms2018performance} as well. 

Machine learning algorithms have significantly contributed to the literature in many studies aimed at separating various BSM signals from backgrounds, not only using CMS or ATLAS data but also using simulated data. In the study \cite{baldi_searching_2014}, researchers developed models to separate signals from backgrounds using various machine learning algorithms, neural networks, and deep learning algorithms for two different beyond the Standard Model signals. One of these signals is the theoretical heavy and neutral Higgs boson production through gluon-gluon fusion, and the other is chargino pair production where these charginos subsequently decay into an LSP, lepton, and neutrino via a W boson. In addition to this study, researchers in \cite{arganda2022towards, celik2023fast} and \cite{celik2024exploring} developed models using CNN, deep learning, transfer learning, and machine learning algorithms to separate the Standard Model background from BSM signals or to isolate 2D histograms representing only the SM background from 2D histograms where the signal is embedded in the SM background. These studies not only compare the performances of the developed models against each other but also demonstrate the successful application of transfer learning in cases with limited data,  particularly as shown in \cite{celik2023fast}.

Recent studies have highlighted the effectiveness of geometric deep learning techniques, especially Graph Neural Networks (GNNs), in tackling various track reconstruction challenges in high-energy particle physics \cite{biscarat2021towards,ju2021performance,dezoort2021charged}. Researchers in the work \cite{dezoort2021charged} have applied a physics-inspired Interaction Network (IN), a specific type of GNN, to address particle tracking under the intense pileup conditions anticipated at the high-luminosity Large Hadron Collider (HL-LHC). By modeling particle tracking data as graphs, where silicon tracker hits serve as nodes and particle trajectories as edges, the IN GNN demonstrated remarkable edge-classification accuracy and tracking efficiency. Notably, the compact IN architecture, significantly smaller than previous GNN tracking models, shows promise for implementation in environments with limited computational resources. The study also investigated the acceleration of IN implementations using diverse computing resources, including FPGAs, and explored alternative GNN approaches that eliminate the need for predefined graph structures. These results emphasize the potential of GNNs, particularly the IN model, as a powerful tool for particle tracking and other reconstruction tasks at the HL-LHC, positioning them as strong candidates for future high-energy physics experiments

The multi-dimensional nature of particle collision data, coupled with the rarity of BSM events, necessitates advanced techniques capable of extracting subtle patterns and features. GNNs offer a promising approach due to their ability to explicitly capture the relationships between particles in each event. This study aims to leverage the strengths of GNNs in handling complex, graph-structured data to improve the sensitivity and efficiency of BSM signal detection. By comparing the performance of GNNs with traditional deep learning methods, this research seeks to provide valuable insights into the most effective approaches for analyzing particle collision data.
Hence, in this study, the classification of gluino-gluino production, a beyond-standard-model signal, from certain backgrounds of the Standard Model in particle physics, using graphical neural networks \cite{4700287}, is addressed. Although GNNs have a wide range of applications, namely in solving many graph-related machine learning problems such as node classification, link prediction, anomaly detection, and graph classification, this study focuses on the latter constructed from tabular data belonging to each class. The performances of the GNNs are then compared with each other and with that of a deep learning method.

\section{Methodology}

The search for BSM signals in particle physics has traditionally relied on conventional methods such as cut-and-count or shape analysis, employed by the researchers. However, the increasing complexity and volume of data generated by modern particle colliders present new challenges that demand more sophisticated analytical approaches. The multi-dimensional nature of particle collision data, coupled with the rarity of BSM events, necessitates advanced techniques capable of extracting subtle patterns and features. In this context, machine learning algorithms, particularly deep learning models, have emerged as powerful tools for analyzing complex datasets in particle physics. This study explores the application of both DNN and GNNs to the challenge of BSM signal detection. By leveraging these advanced machine learning techniques, the aim is to enhance the sensitivity and efficiency of BSM searches, potentially uncovering signals that might elude traditional analysis methods. The comparison between DNNs and GNNs also provides valuable insights into the most effective approaches for handling the unique characteristics of particle collision data, paving the way for more robust and adaptable analysis techniques in high-energy physics.

\subsection{Proposed Method}
This study employs both DNN and GNN models to classify particle collision events, focusing specifically on BSM signal detection against background events. The analysis utilizes background and BSM signal data provided by the research group \cite{darkmachines_community, datas_paper}. The study focused on Channel 1, where the final state consists of Missing Transverse Energy (MET), and jets. The dataset itself includes bunch of different BSM processes such as SUSY and non-SUSY. However, only SUSY gluino-gluino pair production, which eventually decays to jets and neutralinos is considered in this work. Gluinos are assumed to have a mass of 1.6 TeV and the neutralinos have a mass of 800 GeV. The Channel 1, looking for hadronic activity, includes 214000 SM events, and a signal sample of 10755 is constructed as follows: 
\begin{itemize}
    \item $E_{\mathrm{T}}^{\mathrm{miss}} \geq 200 \, \text{GeV}$
    \item $H_T \geq 600 \, \text{GeV}$
    \item $E_{\mathrm{T}}^{\mathrm{miss}} / H_{\mathrm{T}} \geq 0.2 $
    \item $N(j_{p_T > 50 \, \text{GeV}}) \geq 4 $
    \item $N(j_{p_T > 200 \, \text{GeV}}) \geq 1$
\end{itemize}
where j can either be a jet or a b-jet.
Comprehensive details regarding the dataset's production and the pre-selection cuts applied can be found in the referenced article \cite{Aarrestad_2021oeb}.

The model inputs were defined using key physical features of particle collision events: energy (E), transverse momentum ($p_T$), pseudorapidity ($\eta$), and azimuthal angle ($\phi$) for each physics object of interest, including jets, b-jets, electrons, and muons. Additionally, global event features such as MET and its azimuthal angle ($\mathrm{MET}_\phi$) were included. Each event was labeled with a binary target variable (1-0) indicating whether it was a BSM signal (1) or background events (0). However, particle collision events typically contain varying numbers of particles, resulting in inconsistent feature sets across events. To address this variability and ensure consistent input for both the DNN and GNN models, a preprocessing step was implemented. The input features were standardized to include a maximum of four b-jets and four jets per event. For events with more than four objects of a given type, only the first four were retained, while the total count of each object type (e.g., $n_{\mathrm{bjets}}$, $n_{\mathrm{jets}}$) was recorded as an additional feature. For events with fewer than four objects, a feature padding technique was employed by adding placeholder values of $-9999$ for any missing particle data, ensuring a consistent set of features for each event. This approach balances the need for detailed event information with the requirement for uniform input across all events. Both the DNN and GNN models utilize this same preprocessed dataset, allowing for a fair comparison of their performance.

To address class imbalance, undersampling was performed using the pandas library \cite{reback2020pandas}, equalizing the number of signal and background events. For binary classification purposes, all backgrounds contributing to the signal region were treated as a single background class. The resulting dataset was then normalized using the `StandardScaler' class from the scikit-learn library \cite{scikit-learn}, ensuring a mean of 0 and standard deviation of 1 for all features. This normalization step is crucial in preventing bias in machine learning algorithms. Subsequent to normalization, the dataset was split into training, validation, and test sets in a ratio of $60\%$, $20\%$, and $20\%$, respectively.

\begin{figure}[h!tb]
\centering
\includegraphics[width=1.0\textwidth]{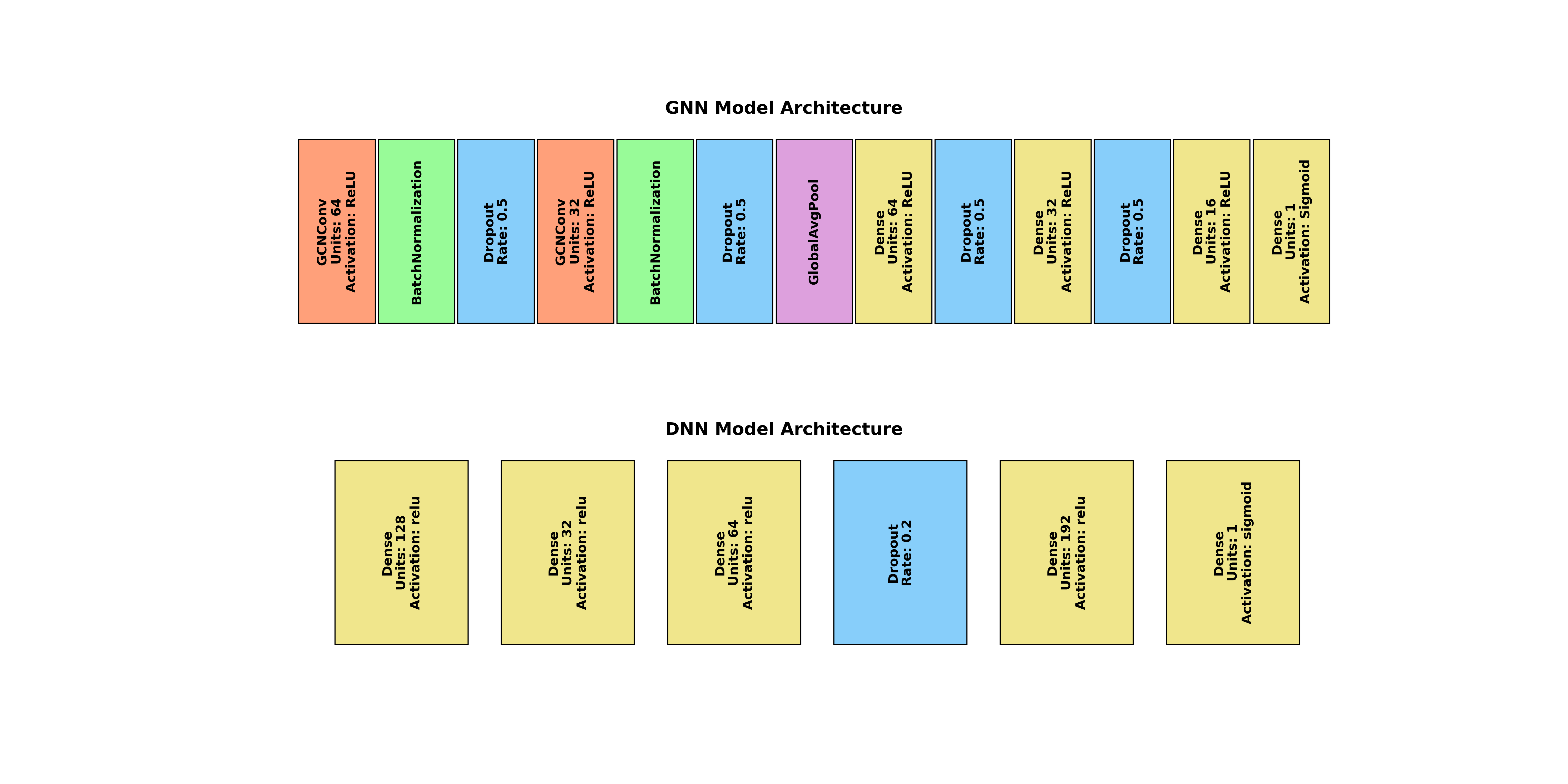}
\caption{Comparison of the architectures of the GNN and DNN models. The GNN model features GCN, Batch Normalization, Dropout, Global Average Pooling, and Dense Layers, While the DNN model includes Dense and Dropout layers.}
\label{model-architecture}
\end{figure}

A DNN model with a flexible architecture was created using Keras API \cite{chollet2015keras}. The model consists of an input layer, four hidden Dense layers with units of 128, 32, 64, and 192, respectively, each employing the ReLU activation function, a Dropout layer with a rate of 0.2 applied after the third Dense layer (64 units) to mitigate overfitting by randomly setting $20\%$ of the input units to zero during training, and an output layer with a sigmoid activation function for binary classification (see Figure \ref{model-architecture} for the architecture of the DNN model). Keras Tuner \cite{omalley2019kerastuner} was used for hyperparameter optimization, searching for the best combination of number of layers, units per layer, dropout rates, and learning rate. The model is compiled using the Adam optimizer with a learning rate of approximately 0.00043 and the binary crossentropy loss function. Training is conducted for a maximum of 100 epochs, with early stopping implemented to halt training when the validation loss ceases to improve, using a patience of 20 epochs. The batch size used during training is 32. To prevent overfitting, early stopping is employed, and when triggered, it ensures that the best model is saved. Hyperparameters that are used and optimized for the final DNN model is listed in Table \ref{tab:hyperparameters}. The final model was created using the best hyperparameters and trained on the training set, validated on the validation set. The model's performance was evaluated on the test set to ensure generalizability. A threshold of 0.5 was applied to the sigmoid output to determine class assignments. Outputs exceeding this threshold were classified as the positive class (1), while those below were assigned to the negative class (0). This consistent threshold was used across all performance metric calculations, including accuracy, precision, recall, and F1 score. The comprehensive training strategy and hyperparameter optimization resulted in a robust DNN model capable of accurately classifying particle collision events based on the provided features.
\begin{table}[h!]
\centering
\caption{Optimum Hyperparameters for GNN and DNN models}
\begin{tabular}{lcc}
\toprule
\textbf{Hyperparameter} & \textbf{GNN} & \textbf{DNN} \\
\midrule
Optimizer & \textit{Adam} & \textit{Adam} \\
Activation function & \textit{ReLu/Sigmoid} & \textit{ReLu/Sigmoid} \\
Loss function & \textit{Binary crossentropy} & \textit{Binary crossentropy} \\
Batch size & 128 & 32 \\
Learning rate & 0.001 & 0.00043 \\

Early Stopping & \textit{yes}  & \textit{yes}\\
Number of Units & 64-32-64-32-16-1 & 128-32-64-192-1 \\
Number of Layers & 13 & 6 \\
\bottomrule
\end{tabular}
\label{tab:hyperparameters}
\end{table}

\begin{figure}[h!tb]
\centering
\includegraphics[width=0.95\textwidth]{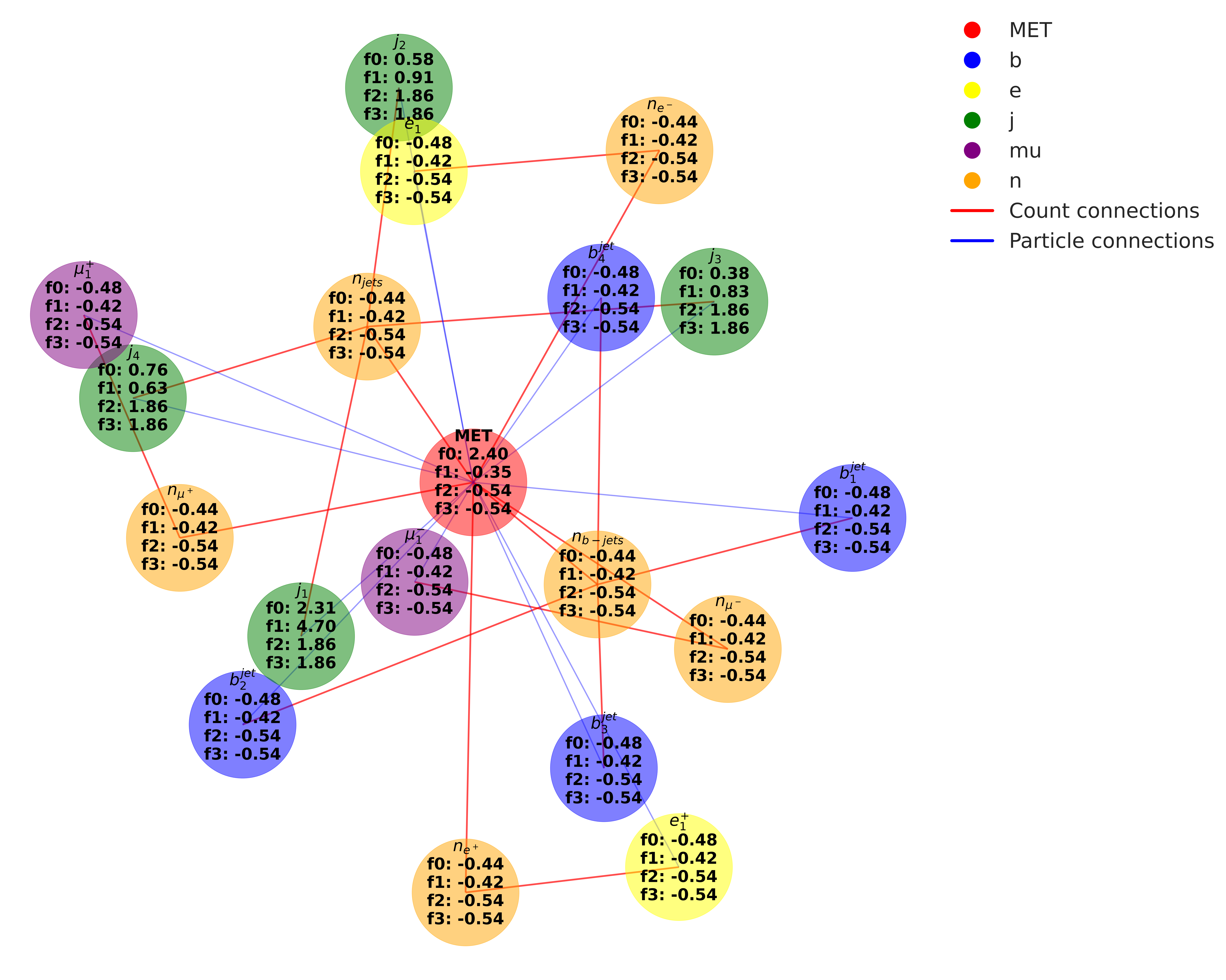}
\caption{Graph representation of the GNN-1 model. The Missing Transverse Energy (MET) node is centrally positioned and connected to all other nodes, forming a star-like topology. This structure emphasizes the global influence of MET on all other particles in the event.}
\label{met-centered}
\end{figure}

\begin{figure}[h!tb]
\centering
\includegraphics[width=1.0\textwidth]{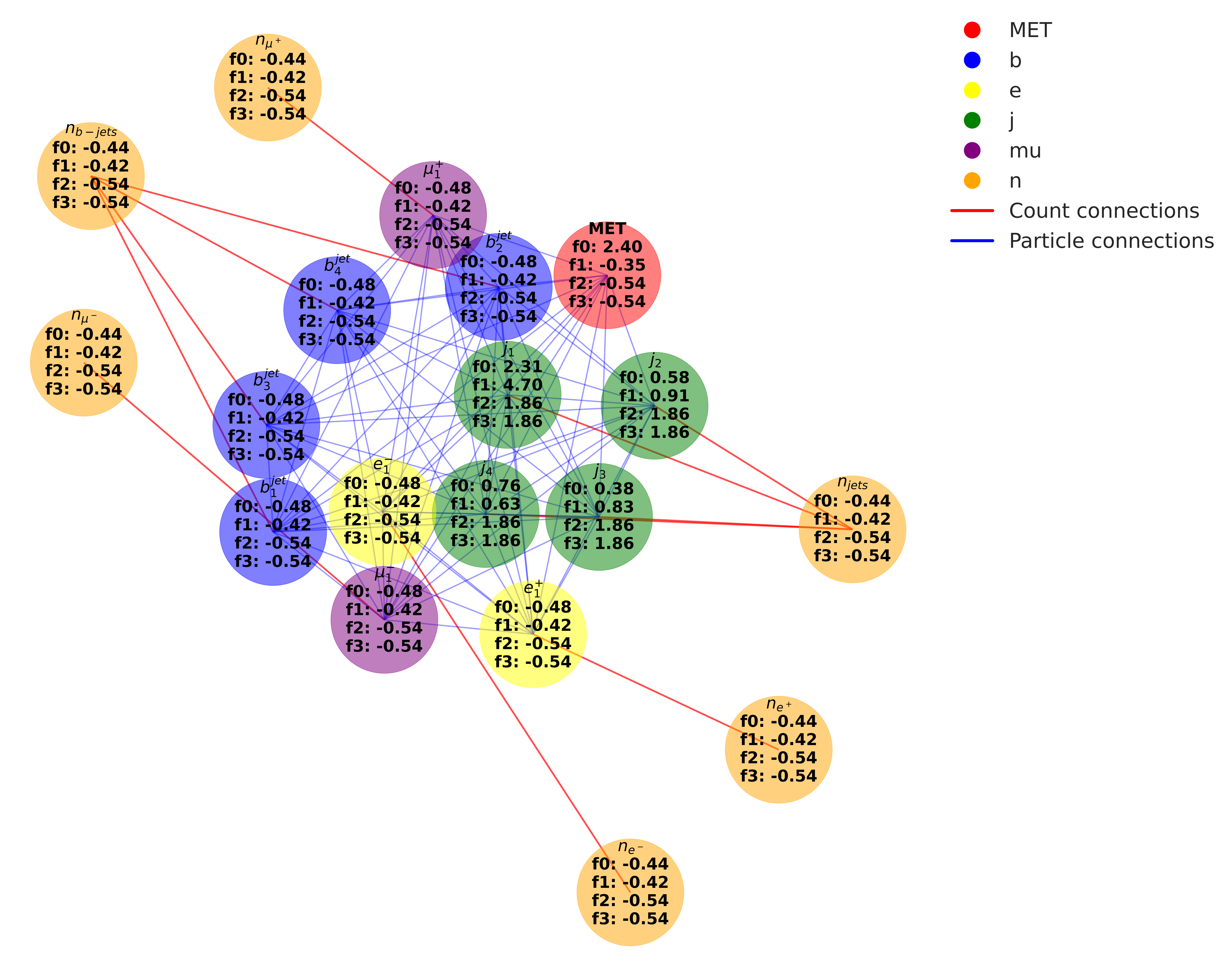}
\caption{Graph representation of the GNN-2 model. The graph features a hybrid connection structure. Most nodes, including MET, are fully interconnected with each other. However, certain nodes maintain specific, physically meaningful connections: the number of electron/positron ($n_{e^{\mp}}$) node links only to the electron/positron ($e^{\mp}$) node; the number of muon/antimuon ($n_{\mu^{\mp}}$) node links only to the muon/antimuon ($\mu^{\mp}$) node; the number of jets ($n_{jets}$) node connects only to jet nodes ($j_1$, $j_2$, $j_3$, $j_4$); and the number of b-jets ($n_{bjets}$) node links only to b-jet nodes ($b_1$, $b_2$, $b_3$, $b_4$).}
\label{all-connected}
\end{figure}

As part of the methodology, two unique graph structures were developed for each event in the dataset. The first structure, named GNN-1, featured MET as the central node, connected to all other feature nodes. The second structure, called GNN-2, was a hybrid connection structure. Most nodes, including MET, are fully interconnected with each other. However, certain nodes maintain specific, physically meaningful connections: the number of electron/positron ($n_{e^{\mp}}$) node links only to the electron/positron ($e^{\mp}$) node; the number of muon/antimuon ($n_{\mu^{\mp}}$) node links only to the muon/antimuon ($\mu^{\mp}$) node; the number of jets ($n_{jets}$) node connects only to jet nodes ($j_1$, $j_2$, $j_3$, $j_4$); and the number of b-jets ($n_{bjets}$) node links only to b-jet nodes ($b_1$, $b_2$, $b_3$, $b_4$).
These graph structures were created using the NetworkX library\cite{hagberg2008}, with node features converted to numpy arrays \cite{harris2020array}. The Spektral library \cite{grattarola2020graph}, which is  based on the Keras API and TensorFlow 2, was then used to adapt these graph structures and features for use in the GNN models. The illustrations of the graphs for both cases are shown in Figure \ref{met-centered} and \ref{all-connected}, respectively.

To effectively manage and prepare the data for training and validation, a custom data loader was implemented. This data loader organized the graph structures into the training, validation, and test sets, ensuring that the data was correctly formatted and optimized for the GNN architecture. By handling the batching and shuffling of the data, the custom loader facilitated efficient training processes and consistent evaluation across the models.

Building on the graph structures developed, two GNN models were implemented to leverage these configurations. The GNN architectures comprise a total of six trainable layers with a sequence of units as 64-32-64-32-16-1. The models, created using Keras and Spektral libraries, consist of input layers, followed by two Graph Convolutional Network layers with 64 and 32 units, respectively, each followed by a batch normalization layer and a dropout layer with a rate of 0.5 to prevent overfitting. After the GCN layers, a global average pooling layer is applied to aggregate the node features. The model then includes three dense layers with 64, 32, and 16 units employing the ReLU activation function. Each of the first two dense layers is followed by a dropout layer with a rate of 0.5. The final output layer is a dense layer with 1 unit and a sigmoid activation function, suitable for binary classification tasks. The architecture of the model is depicted in Figure \ref{model-architecture}. 

Unlike the DNN model, the GNN models were optimized through manual trial and error rather than using tools like Keras Tuner. This approach involved iteratively adjusting the model architecture and hyperparameters based on validation performance. The model was compiled using the Adam optimization algorithm with a learning rate of 0.001 and trained with binary cross-entropy loss. Training was conducted for a maximum of 500 epochs, with early stopping implemented to halt training when the validation loss ceased to improve after 20 consecutive iterations. When early stopping was triggered during the training, the best model was saved. For both GNN models, early stopping and model checkpointing were implemented to prevent overfitting and save the best model during training. The models were trained on the training set and validated on the validation set. The performance of the best model was evaluated on the test set. As with the DNN model, a 0.5 threshold was applied to the GNN outputs for class determination and performance metric calculations, ensuring consistency in evaluation across all models.

 The performance of the two different graph structure types was compared, and the results are presented in Table \ref{tab:model-metrics}. Through meticulous architectural design and manual optimization, the GNN models demonstrated strong performance in classifying particle collision events, highlighting the effectiveness of graph-based approaches in distinguishing BSM signals from background events. 

\subsection{GNN}
A Graph Neural Network is a type of deep learning model designed to work with graph-structured data. Unlike traditional neural networks that operate on grids or sequences, GNNs are capable of handling complex relationships between nodes and edges in a graph. This makes them particularly useful in domains where data is naturally represented as graphs, such as social networks, molecular structures, traffic networks, and recommendation systems. GNNs have been shown to be effective in tasks like node classification, graph classification, and link prediction. 

GNNs are widely used in various applications, including computer vision, natural language processing, and recommender systems \cite{
li2023survey,han2022vision,pradhyumna2021graph,wu2023graph}. For instance, in computer vision, GNNs can analyze the relationships between objects in an image, while in natural language processing, they can model the relationships between words in a sentence. In recommender systems, GNNs can be used to model the relationships between users and items, and generate personalized recommendations. Additionally, GNNs have been applied to various scientific domains, such as chemistry, biology, and physics, to analyze complex systems and make predictions \cite{wang2023graph,reiser2022graph,pata2021mlpf,di2021towards}.

Graph Convolutional Networks are a specific class of neural networks designed to operate on graph-structured data. GCNs can capture the complex relationships and dependencies between nodes in a graph. Mathematically, a GCN extends the concept of convolution from Euclidean space to non-Euclidean space. Given a graph $ G=(V, E) $ with $\mathrm{V}$ set of nodes and $E$ as the set of edges, the graph can be represented by an adjacency matrix  $A$ and a feature matrix $X$. The key operation in a GCN is the graph convolution, which can be expressed as: 

$$ H^{(l+1)}=\sigma\left(\hat{A} H^{(l)} W^{(l)}\right) $$

Here, $ \hat{A}=\tilde{D}^{-1 / 2} \tilde{A} \tilde{D}^{-1 / 2} $ is the normalized adjacency matrix, $ \tilde{D} $ is the degree matrix of  $ \tilde{A}, H^{(l)} $ denotes the feature matrix at layer $ l, W^{(l)} $ is the trainable weight matrix at layer $l$, and $\sigma$ is a non-linear activation function such as ReLU. This operation effectively aggregates the features of neighboring nodes, allowing the network to learn representations that capture the local graph structure.

\subsection{DNN}
Deep Neural Networks represent a significant advancement in machine learning, offering a sophisticated approach to modeling complex systems. Inspired by the neural architecture of the human brain, DNNs extend this concept with multiple hidden layers, enabling them to excel at learning intricate patterns in data.

The architecture of a DNN is characterized by its layered structure, where each successive layer processes and refines the output from the previous one. This hierarchical processing allows DNNs to extract increasingly abstract features or representations from the input data, facilitating the modeling of non-linear relationships that often challenge traditional algorithms.

The versatility of DNNs has led to their widespread adoption across various domains. In computer vision, they have revolutionized tasks such as object recognition, image classification, and even image synthesis. Similarly, the field of natural language processing has seen remarkable progress through the application of DNNs in areas like machine translation, sentiment analysis, and conversational AI. 

Moreover, the impact of DNNs extends beyond these computational fields. In healthcare, they are instrumental in advancing diagnostic techniques and accelerating drug discovery processes. The financial sector leverages DNNs for fraud detection and risk assessment, demonstrating their utility in handling complex, multidimensional data. 

The key strength of DNNs lies in their ability to distill meaningful insights from large, complex datasets. This capability positions them as invaluable tools in our increasingly data-driven world, where the extraction of actionable information from vast amounts of data is crucial for advancement in numerous fields of study and industry.

\section{Findings}

In this study, a total of three models, two of which are GNNs, were developed, and their performances are presented comparatively. In both GNN-1 and GNN-2 models, each simulated event corresponds to a graph, with each particle or, in other words, physics object corresponding to a node, and attributes related to this objects are referred as features. Since missing transverse momentum is an important parameter in BSM research, in the first GNN model created, graphs were constructed with the MET node at the center and connected to other nodes, while in the second type, a model was created by selecting all nodes to be interconnected. Looking at the results of both of these models, it is clear that they provide similar performance across all the metrics.

\begin{figure}[ht]
\centering
\includegraphics[width=0.95\textwidth]{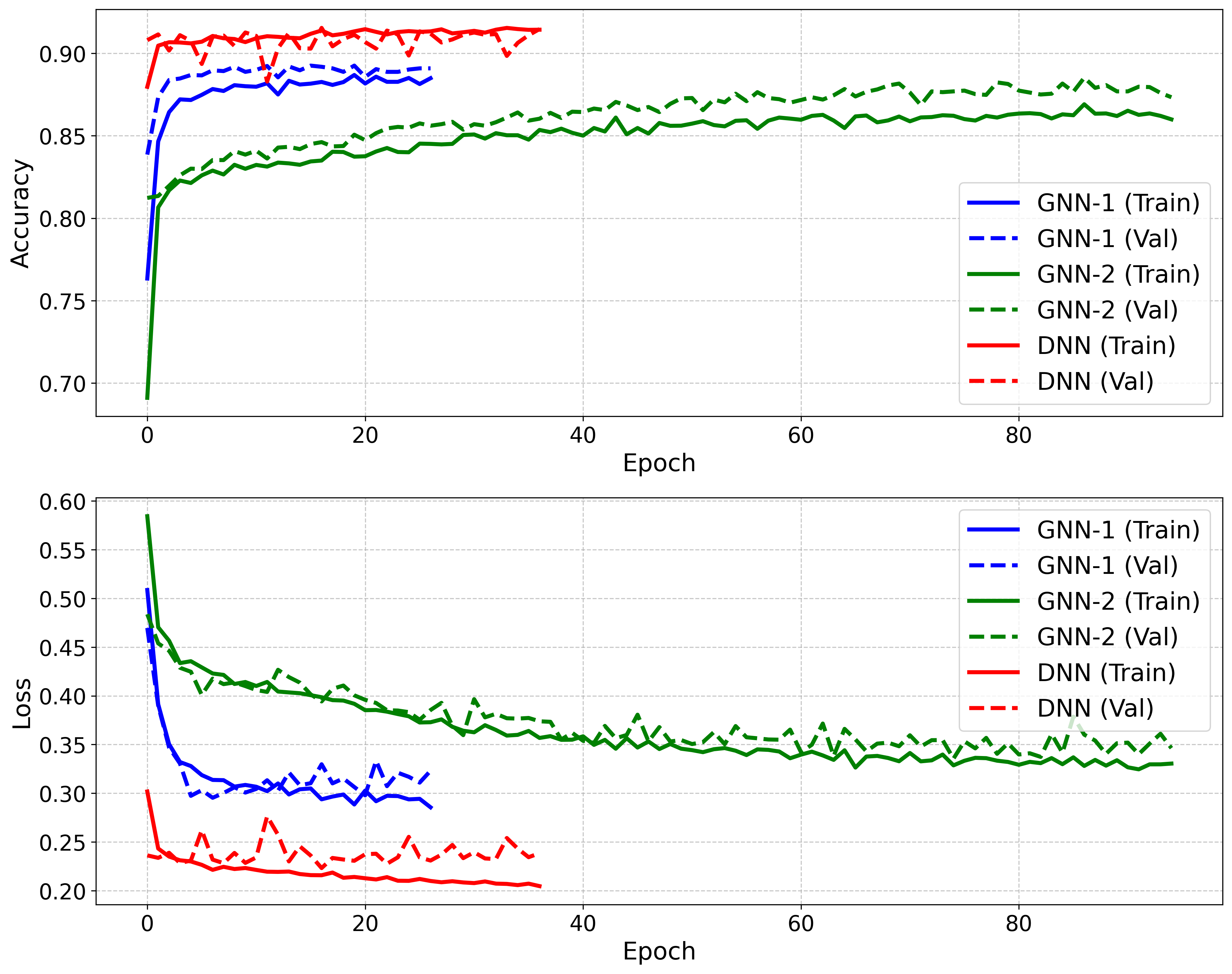}
\caption{Comparison of model performance: Top - Accuracy of GNN-1, GNN-2, and DNN models on training and validation data; bottom - training and validation loss over epochs for all three models. }
\label{accuracy-and-loss}
\end{figure}

\begin{table}[ht]
\centering

\resizebox{\textwidth}{!}{%
\begin{talltblr}[caption={Confusion Matrices for GNN-1, GNN-2, and DNN Models in Classifying Background and Signal Events}, label = {tab:confussion_matrix_table}]{
  row{1} = {c},
  row{2} = {c},
  row{3} = {c},
  row{4} = {c},
  row{5} = {c},
  cell{1}{1} = {r=5}{},
  cell{1}{4} = {c=2}{},
  cell{1}{7} = {r=5}{},
  cell{1}{10} = {c=2}{},
  cell{3}{2} = {r=2}{},
  cell{3}{8} = {r=2}{},
  cell{6}{4} = {r=5}{c},
  cell{6}{5} = {c},
  cell{6}{6} = {c},
  cell{6}{7} = {c=2}{c},
  cell{6}{9} = {c},
  cell{6}{10} = {c},
  cell{6}{11} = {c},
  cell{6}{12} = {c},
  cell{7}{5} = {c},
  cell{7}{6} = {c},
  cell{7}{7} = {c},
  cell{7}{8} = {c},
  cell{7}{9} = {c},
  cell{7}{10} = {c},
  cell{7}{11} = {c},
  cell{8}{5} = {r=2}{c},
  cell{8}{6} = {c},
  cell{8}{7} = {c},
  cell{8}{8} = {c},
  cell{8}{9} = {c},
  cell{8}{10} = {c},
  cell{8}{11} = {c},
  cell{9}{6} = {c},
  cell{9}{7} = {c},
  cell{9}{8} = {c},
  cell{9}{9} = {c},
  cell{9}{10} = {c},
  cell{9}{11} = {c},
  cell{10}{5} = {c},
  cell{10}{6} = {c},
  cell{10}{7} = {c},
  cell{10}{8} = {c},
  cell{10}{9} = {c},
  cell{10}{10} = {c},
  cell{10}{11} = {c},
  vline{1-2,7-8,13} = {1-5}{},
  vline{7,13} = {2,5}{},
  vline{3-7,9-13} = {3-4}{},
  vline{4-7,10-13} = {4}{},
  vline{4-5,10} = {6-10}{},
  vline{4,10} = {7,10}{},
  vline{4,6-10} = {8-9}{},
  vline{4,7-10} = {9}{},
  hline{1,6} = {-}{},
  hline{3-5} = {3-6,9-12}{},
  hline{8-10} = {6-9}{},
  hline{11} = {4-9}{},
}
\textbf{\begin{sideways}GNN-1\end{sideways}} &        &                 & \textit{Predicted} &                 &                 & \textbf{\begin{sideways}GNN-2\end{sideways}}              &                 &                 & \textit{Predicted} &                 &                \\
      &        &                 & \textbf{BG's}      & \textbf{Signal} & \textbf{Total}  &                    &                 &                 & \textbf{BG's}      & \textbf{Signal} & \textbf{Total} \\
      & \textit{\begin{sideways}Actual\end{sideways}} & \textbf{BG's}   & 1896               & 252             & 2148            &                    & \textit{\begin{sideways}Actual\end{sideways}}          & \textbf{BG's}   & 1896               & 252             & 2148           \\
      &        & \textbf{Signal} & 233                & 1921            & 2154            &                    &                 & \textbf{Signal} & 279                & 1875            & 2154           \\
      &        & \textbf{Total:} & 2129               & 2173            & 4302            &                    &                 & \textbf{Total:} & 2175               & 2127            & 4302           \\
      &        &                 & \textbf{\begin{sideways}DNN\end{sideways}}                &                 &                 & \textit{Predicted} &                 &                 &                    &                 &                \\
      &        &                 &                    &                 &                 & \textbf{BG's}      & \textbf{Signal} & \textbf{Total}  &                    &                 &                \\
      &        &                 &                    & \textit{\begin{sideways}Actual\end{sideways}}        & \textbf{BG's}   & 1960               & 188             & 2148            &                    &                 &                \\
      &        &                 &                    &                 & \textbf{Signal} & 200                & 1954            & 2154            &                    &                 &                \\
      &        &                 &                    &                 & \textbf{Total:} & 2160               & 2142            & 4302            &                    &                 &                

\end{talltblr}}
\end{table}
Examining the loss and accuracy plots in Figure \ref{accuracy-and-loss}, both training and validation loss follow similar trends for all three models, and training was terminated with early stopping as there was no improvement in validation loss after a certain epoch. It is also worth noting that, based on these loss graphs, overfitting does not appear to be an issue for any of the models. Additionally, as shown in Table \ref{tab:model-metrics}, the DNN demonstrates slightly better results in terms of both accuracy and other measures. However, no significant difference was observed when compared to the GNN models.

In addition to Recall, Precision, Accuracy, F1, the AUC was also used as a performance metric in this study. AUC is a metric that takes values between 0.5 and 1.0 and is defined as the area under the true positive rate (TPR) vs false positive rate (FPR) graph. While a value of 0.5 indicates performance no better than random guessing, an AUC value of 1.0 suggests that the model is an excellent classifier. A value of 0.7 can be interpreted as the model being successful. As shown in Table \ref{tab:model-metrics}, all three models, DNN, GNN-1, and GNN-2, demonstrated excellent performance, with AUC values exceeding $94\%$, with the DNN achieving slightly better performance. 

To further evaluate the models' performance, confusion matrices were generated for each of the three models (see Table \ref{tab:confussion_matrix_table}). These matrices provide a detailed breakdown of true positives, true negatives, false positives, and false negatives, offering insight into the specific strengths and weaknesses of each model in classifying signal and background events.

\begin{figure}[ht]
\centering
\includegraphics[width=0.80\textwidth]{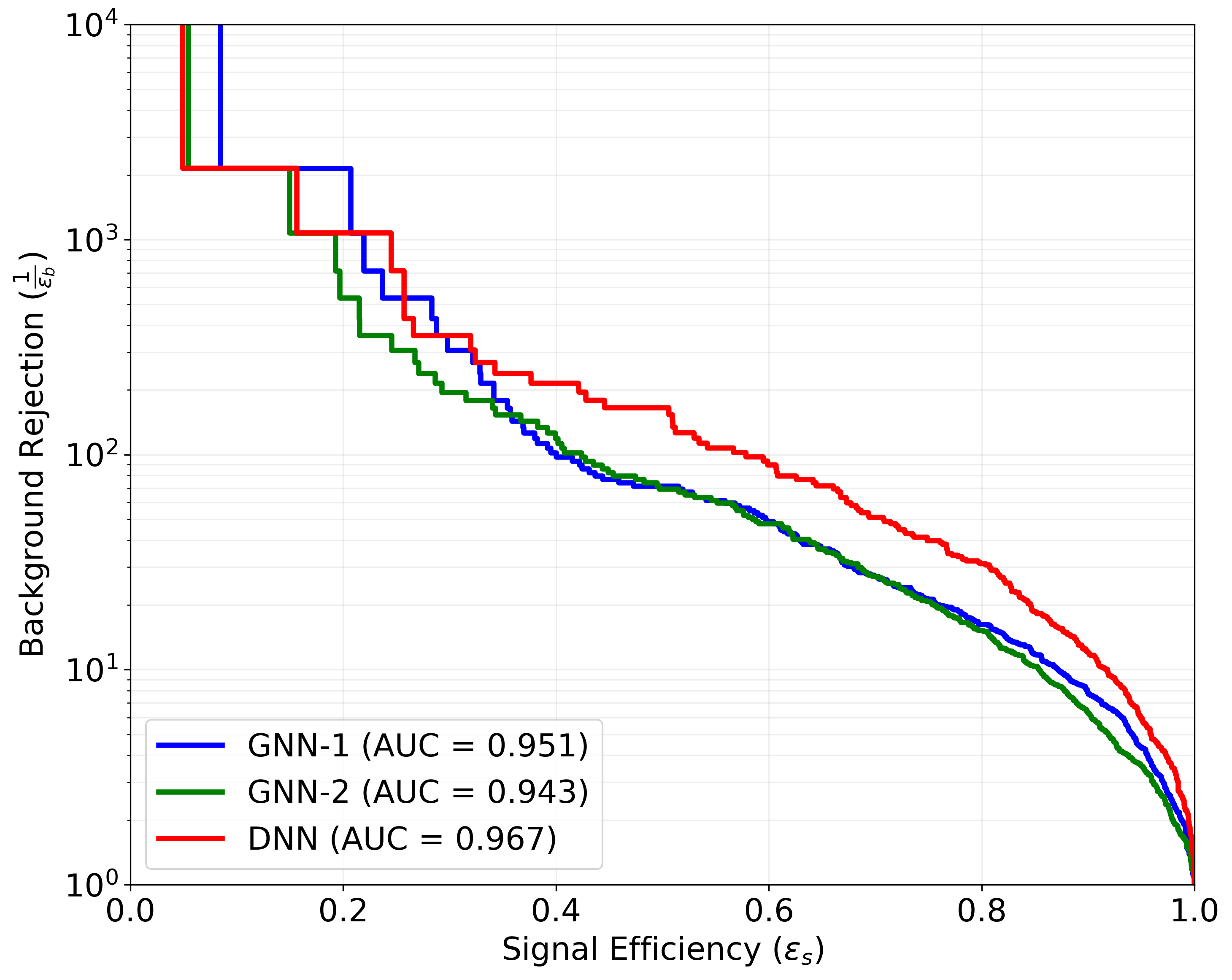}
\caption{Comparison of model performance: ROC curves showing signal efficiency vs background rejection for two GNN's (GNN-1, GNN-2) and a DNN. DNN shows the highest Area Under the Curve (AUC) of 0.967, outperforming both GNN models, particularly beyond signal efficiency of 0.4 where it achieves slightly higher background rejection. GNN-1 (AUC = 0.951) slightly outperforms GNN-2 (AUC = 0.943). All models converge at very high signal efficiencies and show similar trends at low signal efficiency regions, indicating comparable performance in both high-recall and high-precision scenarios.}
\label{fig:roc-combined}
\end{figure}

Figure \ref{fig:roc-combined} provides further insight into the models' performance by plotting the background rejection as a function of signal efficiency. This representation, often referred to as a ROC curve in particle physics contexts, allows for a more nuanced comparison of the models' discriminative power across different operating points. Notably, all three models exhibit strong performance. The DNN model shows slightly better performance, with its curve slightly higher than the GNN variants across most of the signal efficiency range. This is consistent with its higher AUC value. The similarity in performance between the DNN and GNN models suggests that both approaches are capable of effectively capturing the complex relationships in particle collision data. The strong performance of the GNN models indicates that the graph-based representation of collision events preserves important topological information, which can be effectively leveraged for signal-background discrimination.
Notably, the models achieved high classification accuracy using only low-level features, without relying on high-level engineered variables. This finding suggests that even without high-level features, the models can effectively capture the necessary information for accurate classification. This represents a significant advantage in simplifying data preparation and potentially reducing the need for extensive
feature engineering and analysis.

Beyond the statistical performance, these results have important physics implications. The ability of both DNN and GNN models to accurately distinguish between SM and BSM events suggests that such machine learning techniques can enhance the sensitivity of BSM searches. By effectively identifying rare signal events, these models could contribute to refining current experimental limits on BSM theories, potentially leading to the discovery of new physics phenomena.

\section{Conclusion}

This study has demonstrated the efficacy of both DNN and GNNs in classifying particle collision events, with a focus on detecting Beyond Standard Model signals against background events. The comparative analysis of these models offers valuable insights into their performance and potential applications in high-energy physics research.

The strong performance of the GNN models, despite different graph construction methods, highlights the potential of graph-based representations in preserving crucial topological information about particle collisions. The application of GNNs presents a novel approach to particle physics analyses, reflecting the data structure more accurately by modeling the interactions between particles as graphs. This allows GNNs to capture complex dependencies and relational information inherent in collision events, which are often challenging to address with traditional methods.

By leveraging the inherent connections in the data, GNNs offer a unique advantage in detecting subtle patterns and correlations that may be indicative of new physics phenomena. This suggests that GNNs could be a valuable tool in future particle physics analyses, offering a complementary approach to traditional methods. It's worth noting that while DNN model provided slightly better results in this study, the performance of GNNs could potentially be enhanced with access to larger datasets and further optimization, underscoring their promise in this field for advancing BSM searches.

The study's methodology, including undersampling to address class imbalance and careful feature engineering, proved effective in creating models that generalize well to unseen data. The high recall and precision scores across all models indicate their ability to accurately identify rare signal events, a crucial capability in BSM physics research.
\begin{table}[htbp]
    \centering
    \caption{Comparison of performance metrics for GNN and DNN models. The metrics include Accuracy, Recall, Precision, F1 Score, and Area Under the Curve}
    \label{tab:model-metrics}
    \begin{tabular}{lcccccc}  
        \toprule
        \textbf{Model} & \textbf{Accuracy} & \textbf{Precision} & \textbf{Recall} & \textbf{F1 Score} & \textbf{AUC} \\
        \midrule
        GNN-1 & $88.7\%$ & $88.4\%$ &$89.2\%$ & $88.8\%$ & $95.1\%$ \\
        GNN-2 & $87.7\%$ & $88.2\%$ & $87.1\%$ & $87.7\%$ & $94.3\%$ \\
        DNN & $91.0\%$ & $91.2\%$ & $90.7\%$ & $91.0\%$ & $96.7\%$ \\
        \bottomrule
    \end{tabular}
    
\end{table}
While this study focused on gluino-gluino pair production, the strong performance of these models suggests potential applicability to other BSM processes. This versatility, coupled with the rapid advancements in machine learning, positions these techniques at the forefront of high-energy physics research. As the field continues to evolve, deep learning approaches like those explored in this study are poised to play an increasingly important role in pushing the boundaries of our understanding.

Building on these promising results, future research should explore the scalability and adaptability of these models. Particularly, incorporating GNNs into larger, more representative datasets typical in full-scale particle physics experiments could further enhance their effectiveness. Such advancements may contribute significantly to refining the current limits of BSM searches and potentially uncovering new physics. This study underscores the potential of deep learning approaches in particle physics analysis and motivates further investigation of GNN architectures for BSM signal detection.

\bibliographystyle{IEEEtran}  
\bibliography{bibtex}     
\end{document}